\documentclass[10pt,aps,pre,twocolumn,superscriptaddress,showpacs,preprintnumbers]{revtex4-1}

\usepackage{amsmath,amssymb,amsfonts}
\usepackage{graphicx,color}
\usepackage{dcolumn}
\usepackage{bm}
\usepackage{hyperref}

\newcommand{\avg}[1]{\langle #1\rangle}
\newcommand{\ie}{\emph{i.e.}}

\begin{document}

\title{Numerical assessment of the percolation threshold using complement networks} 

\author{Giacomo Rapisardi}
\email{giacomo.rapisardi@imtlucca.it}
\affiliation{IMT School for Advanced Studies, 55100 - Lucca, Italy}
\author{Guido Caldarelli}
\affiliation{IMT School for Advanced Studies, 55100 - Lucca, Italy}
\affiliation{Istituto dei Sistemi Complessi (ISC)-CNR, 00185 - Rome, Italy}
\author{Giulio Cimini}
\affiliation{IMT School for Advanced Studies, 55100 - Lucca, Italy}
\affiliation{Istituto dei Sistemi Complessi (ISC)-CNR, 00185 - Rome, Italy}
\date{\today}

\begin{abstract}
Models of percolation processes on networks currently assume locally tree-like structures at low densities, and are derived exactly only in the thermodynamic limit. 
Finite size effects and the presence of short loops in real systems however cause a deviation between the empirical percolation threshold $p_c$ and its model-predicted value $\pi_c$. 
Here we show the existence of an empirical linear relation between $p_c$ and $\pi_c$ across a large number of real and model networks.
Such a putatively universal relation can then be used to correct the estimated value of $\pi_c$. We further show how to obtain a more precise relation using the concept of the complement graph, 
by investigating on the connection between the percolation threshold of a network, $p_c$, and that of its complement, $\bar{p}_c$.
\end{abstract}

\maketitle 


Percolation theory is a widely used concept in statistical physics \cite{Stauffer1994}, 
in particular in the field of complex networks to study critical phenomena, 
resilience and spreading processes \cite{Callaway2000,Newman2002,Dorogovtsev2008}. 
However, percolation properties in network models (be they sparse treelike graphs \cite{Cohen2002} 
or clustered networks \cite{Serrano2006,Newman2009}) are often considerably 
different from those of real-world networks---which feature a highly more complex topology. 
Recently, percolation has been reformulated as a message passing process 
which takes as input the detailed topology of a given network to predict 
percolation-related observables \cite{Karrer2014,Hamilton2014}, and which implies 
that the bond percolation threshold $\pi_c$ of the network is bounded from below 
by the leading eigenvalue of its non-backtracking matrix \cite{Hashimoto2014}. 
This approach has been then generalized to clustered networks \cite{Radicchi2016}, 
in order to go beyond the locally treelike approximation which is not reliable 
for networks with high density of edges and short loops \cite{Radicchi2015b}. 
However, the method comes at a price of much higher computational complexity, 
and is not particularly satisfactory for bond percolation processes. 
Another important aspect of the message passing approach is that it describes network percolation in the thermodynamic limit, 
and as such cannot be precisely applied to finite graphs \cite{Timar2017}. 
Indeed, the very percolation transition is ill defined in finite systems. 

Numerical simulations of the percolation process obtain the value of the percolation threshold $p_c$ using Monte Carlo techniques \cite{Newman2000}. 
Given $Q$ independent realizations of the process at fixed percolation probability $p$, 
and the relative size of the largest cluster in the network $s_q(p)$, $q=1,\dots,Q$ in the $q$-th realization, 
the percolation strength at $p$ is estimated as $S(p)=\tfrac{1}{Q}\sum_qs_q(p)$, 
and the susceptibility as $\chi(p)=\tfrac{1}{Q\cdot S(p)}\sum_q[s_q(p)]^2-S(p)$.
The best estimate of the percolation threshold is then the value of $p$ at which the susceptibility is maximal.
As the simulated system is finite, such defined pseudo-critical threshold $p_c(N)$ decays towards the percolation threshold 
as $p_c(N)-p_c\sim N^{-1/\nu}$, where $N$ is the (finite) size of the network \cite{Radicchi2015a}.

\begin{figure}
\centering
\includegraphics[scale=.42]{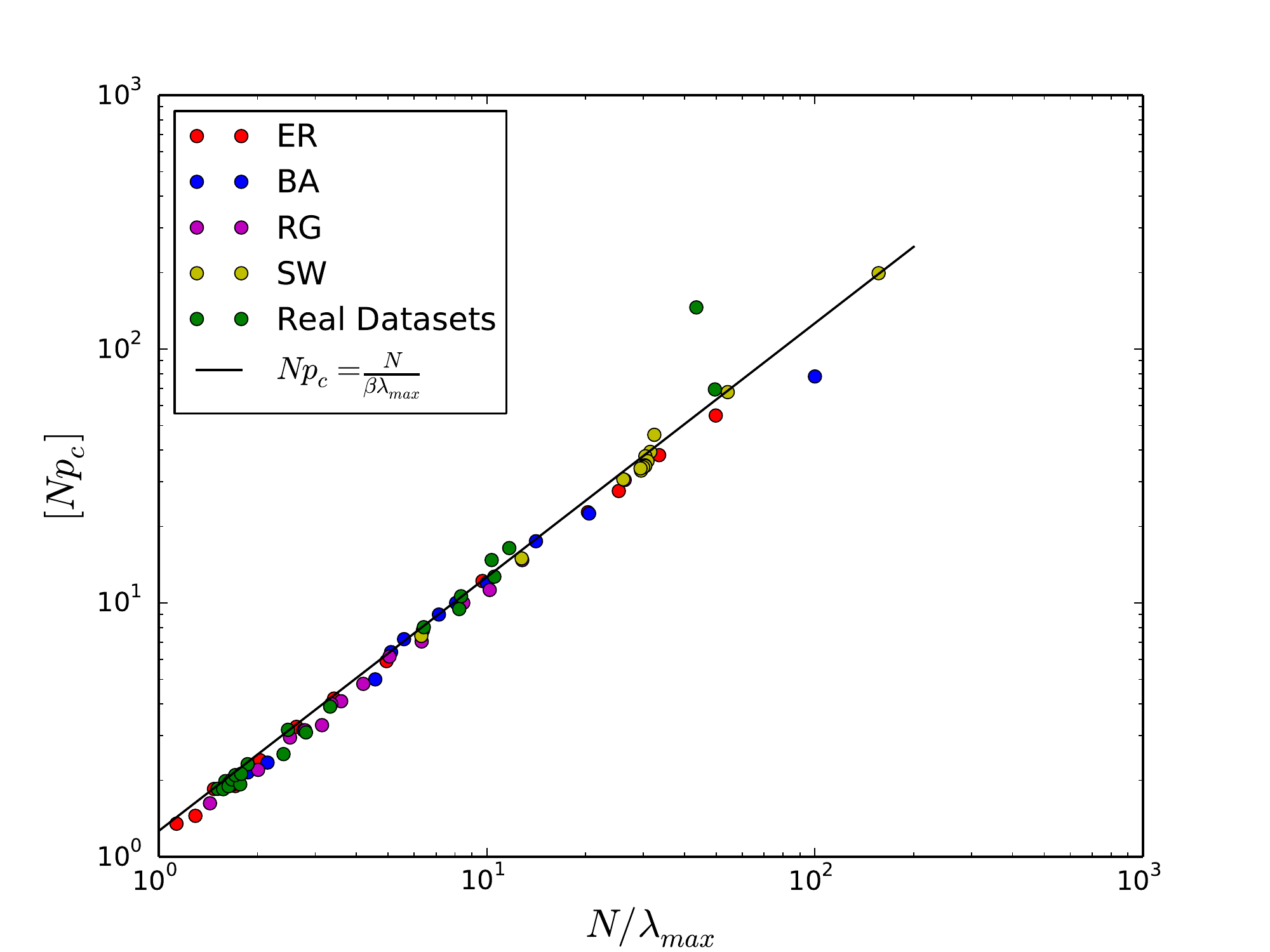}
\caption{Plot of $Np_c$ versus $N\pi_c=N/\lambda_{max}$, where $\lambda_{max}$ is the leading eigenvalue of the non-backtracking matrix, for several model and empirical networks. 
Note that accounting for the factor $N$ allows to compare networks of different size. The black solid line is the linear fit $p_c=\pi_c/\beta$.}\label{fig:compare_p}
\end{figure}

Figure \ref{fig:compare_p} shows, for the bond percolation problem, the relation between the value $p_c$ obtained in numerical simulations 
and the theoretical $\pi_c$ given by the inverse of the leading eigenvalue of the non-backtracking matrix ($\lambda_{max}$). 
The plot is obtained by considering a total of 79 networks of different sizes $N$ (varying approximately from 20 to 890), 23 of which are empirical 
while the remaining 56 are artificially generated according to four different random network models: 
Erd\"os-R\'enyi (ER), Regular (RG), Barab\'asi-Albert (BA) and Watts-Strogatz (SW) \cite{Newman2003}. 
Points are well fitted by a linear relation $p_c=\pi_c/\beta$ with $\chi^2/\nu=4.34$, where however the value of $\beta=0.791\pm0.019$ is different from unity: 
numerical and theoretical percolation thresholds do not coincide, yet their ratio appears to be constant across a variety of empirical and model networks of different size. 
While assessing the general validity of such an empirical evidence needs further statistical analysis, 
this relation can be quite valuable for correcting the theoretical value of $\pi_c$ for finite, non-treelike networks. 

In this work we explore the possibility to improve such an empirical relation using the concept of {\em complement graph}. 
The complement of a graph $G$ is the graph $\bar{G}$ with the same vertex set, but whose edges are those which are not present in $G$ \cite{Clark1983,Gross1998}. 
The union graph of $G$ and $\bar{G}$ is therefore a complete graph. Complement graphs are found since long in the mathematic literature, 
for instance to address the graph coloring problem \cite{Nordhaus1956}, to develop graph compression schemes \cite{Kao1998} and search algorithms \cite{Ito1998}, 
to study network synchronizability \cite{Duan2008}, to assess graph hyperbolicity \cite{Bermudo2011} and domination numbers \cite{Haas2004}. 
The common approach of these studies is to prove rigorous results for graphs with a small number of vertices \cite{Akiyama1979,Xu1987,Petrovic2003}. 
Here, for the first time to our knowledge, we use complement graphs in the context of percolation on large-scale complex networks. 
In particular, we investigate on the existence of a complement relation for the percolation threshold $p_c$ of a given graph $G$ 
and the complement percolation threshold $\bar{p}_c$ of $\bar{G}$.

\begin{figure}
\includegraphics[scale=.42]{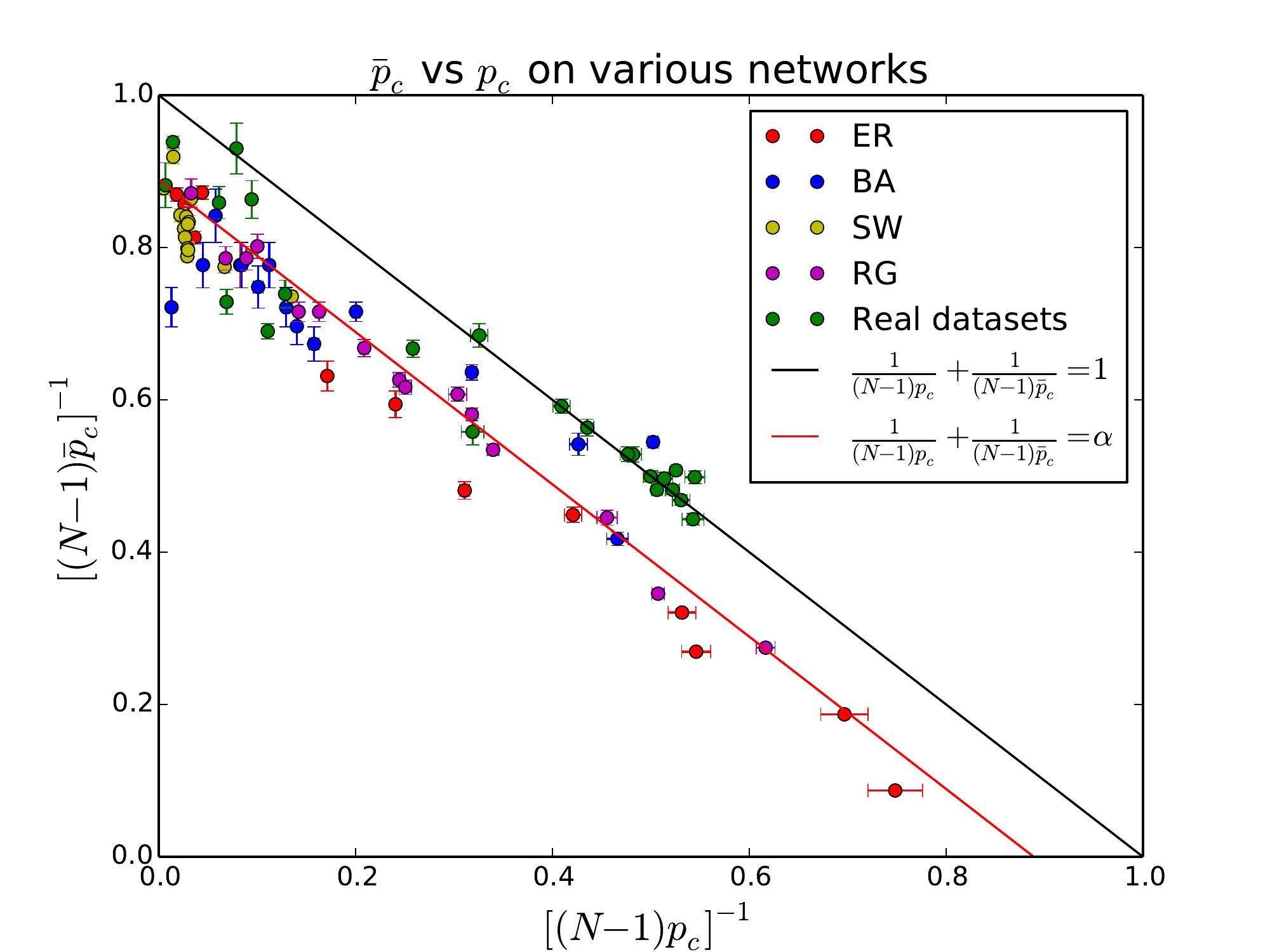}
\caption{Plot of $p_c^{-1}$ versus $\bar{p}_c^{-1}$ for various networks of different sizes. 
The solid black line is eq. \eqref{eqn:p_c}, while the solid red line is the linear fit of data.}\label{fig:p_vs_cp}
\end{figure}

\begin{figure}
\centering
\includegraphics[scale=.42]{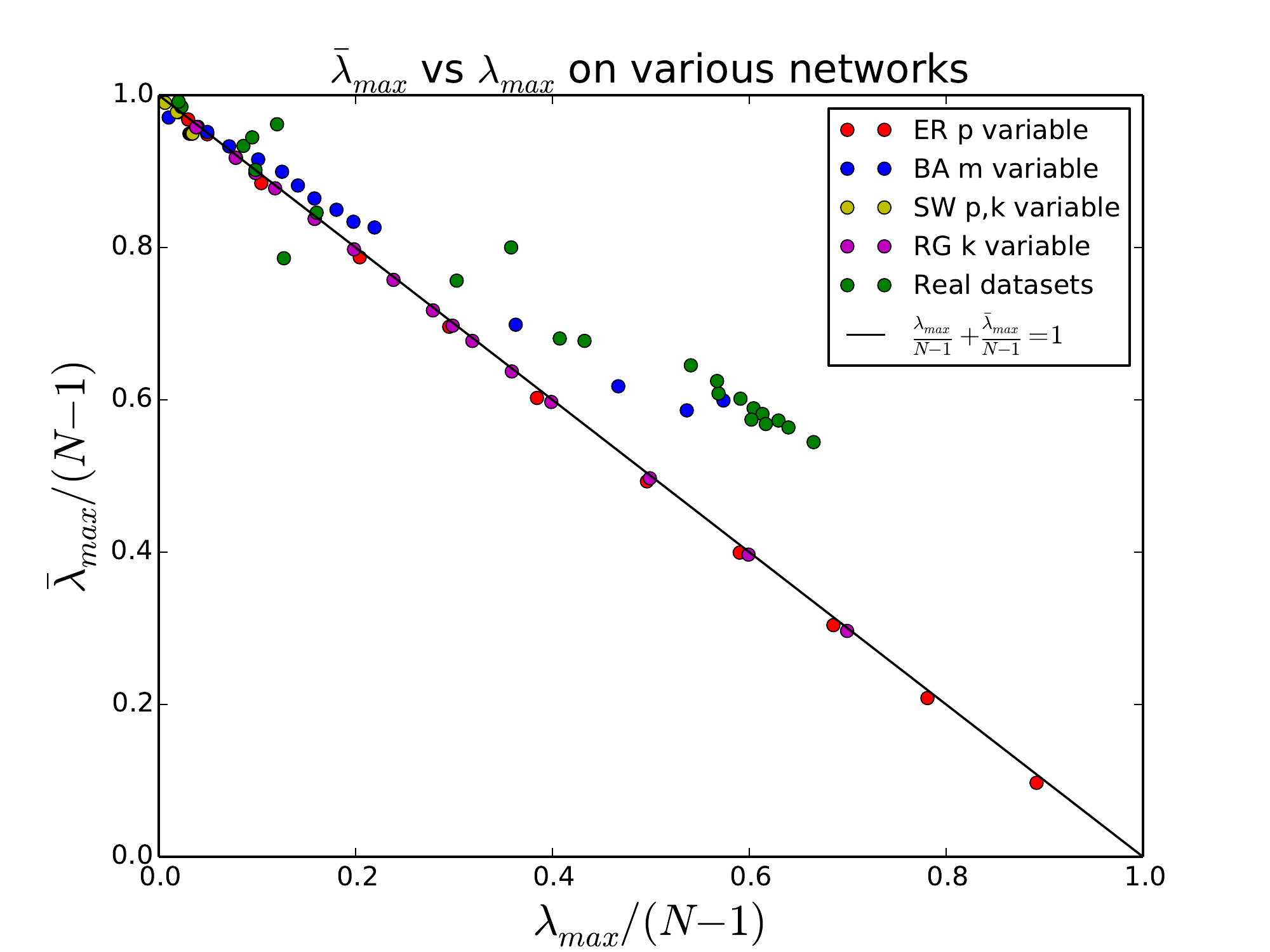}
\caption{Plot of $\lambda_{max}$ versus $\bar{\lambda}_{max}$ for various networks of different sizes.}\label{fig:l_vs_cl}
\end{figure}

Now, since the complement of a sparse network is dense, in the thermodynamic limit the percolation threshold of $\bar{G}$ 
converges to the inverse of the leading eigenvalue of the adjacency matrix of $\bar{G}$ \cite{Bollobas2010}. 
In the simple case of ER networks, for $N\to\infty$ it is $p_c\simeq 1/\avg{k}=1/[(N-1)f]$ (where $f$ is the probability of existence of an edge), 
and thus the following relation should hold:
\begin{equation}\label{eqn:p_c}
\frac{1}{(N-1)p_c}+\frac{1}{(N-1)\bar{p}_c}\simeq 1
\end{equation}
(an analogous complement relation of the two critical points also holds for regular graphs). 
As Figure \ref{fig:p_vs_cp} shows, eq. \eqref{eqn:p_c} slightly overestimates the relation between $p_c$ and $\bar{p}_c$, as they do not add up to unity. 
In particular, the theoretical curve seems to constitute a boundary in the $(p_c,\bar{p}_c)$ plane, 
and data are better fitted by a shifted linear relation
\begin{equation}\label{eqn:p_c_true}
\frac{1}{(N-1)p_c}+\frac{1}{(N-1)\bar{p}_c}=\alpha<1, 
\end{equation}
with $\alpha=0.889\pm0.008$ and $\chi^2/\nu=3.68$. 
The same behavior is observed in Figure \ref{fig:l_vs_cl} for theoretical values of the percolation threshold, obtained as the leading eigenvalue of the non-backtracking matrices.

\begin{figure}
\centering
\includegraphics[scale=.42]{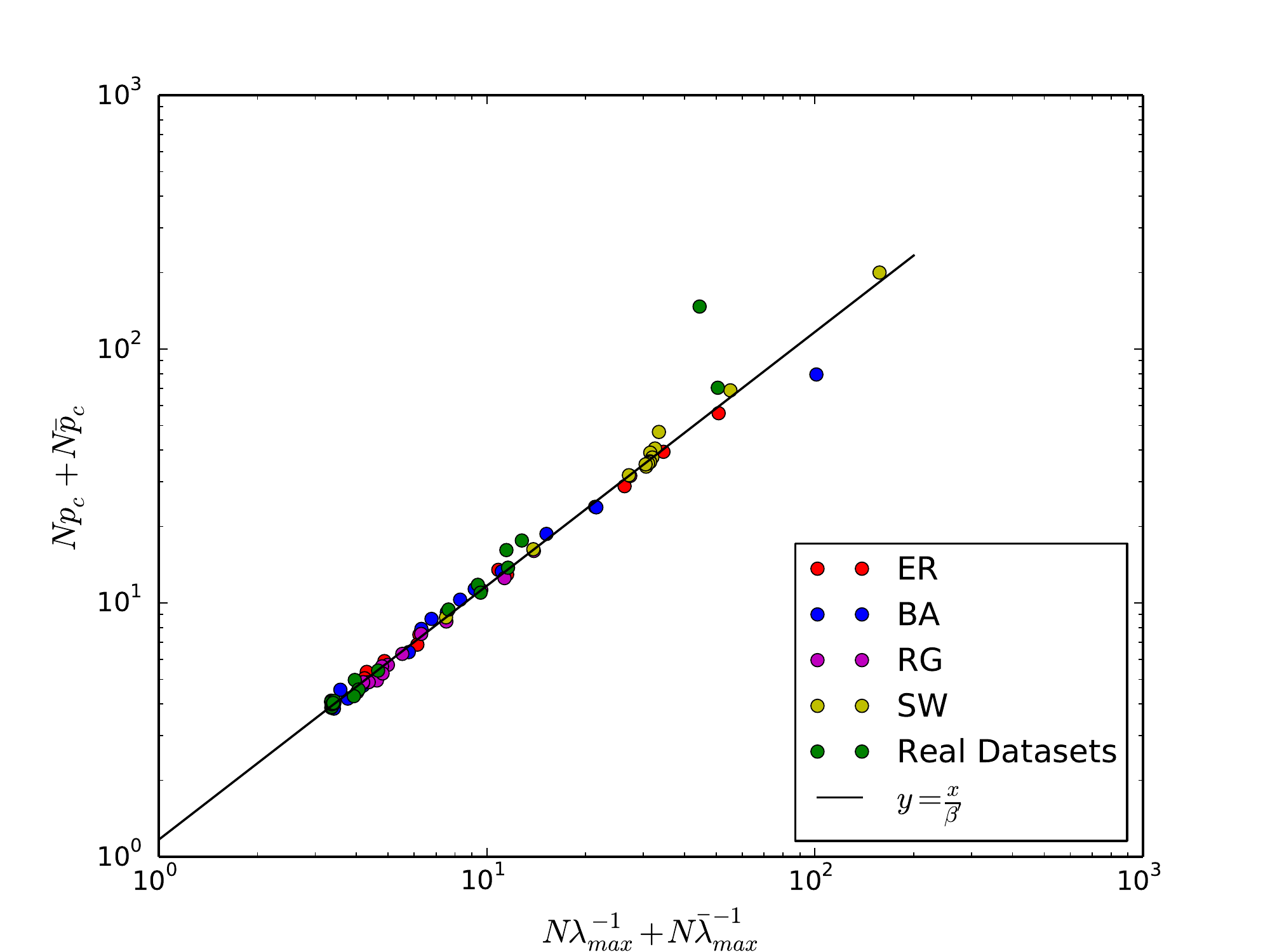}
\caption{Plot of $Np_c + N\bar{p}_c$ vs $\Big( \frac{N}{\lambda_{max}} + \frac{N}{\bar{\lambda}_{max}}\Big)$, 
where $\lambda_{max}$ is the leading eigenvalue of the non-backtracking matrix, for several model and empirical networks. 
The black solid line is the linear fit of eq. \eqref{eqn:lxly_vs_pxpy}.}\label{fig:compare_p-cp}
\end{figure}

Building on the analysis of Figure \ref{fig:compare_p}, we now study the relation
\begin{equation}\label{eqn:lxly_vs_pxpy}
p_c + \bar{p}_c = \frac{1}{\beta'}\left( \frac{1}{\lambda_{max}} + \frac{1}{\bar{\lambda}_{max}}\right).
\end{equation}
As shown in Figure \ref{fig:compare_p-cp}, eqn.(\ref{eqn:lxly_vs_pxpy}) fits the data quite well, 
and much better than the fit of Figure \ref{fig:compare_p}. From the fit we obtained $\beta' = 0.856 \pm 0.010$ and $\chi^2/\nu=1.16$.
This factor can therefore be used to improve the estimate of the percolation threshold on finite, non treelike networks. 

\begin{figure}
\centering
\includegraphics[scale=.42]{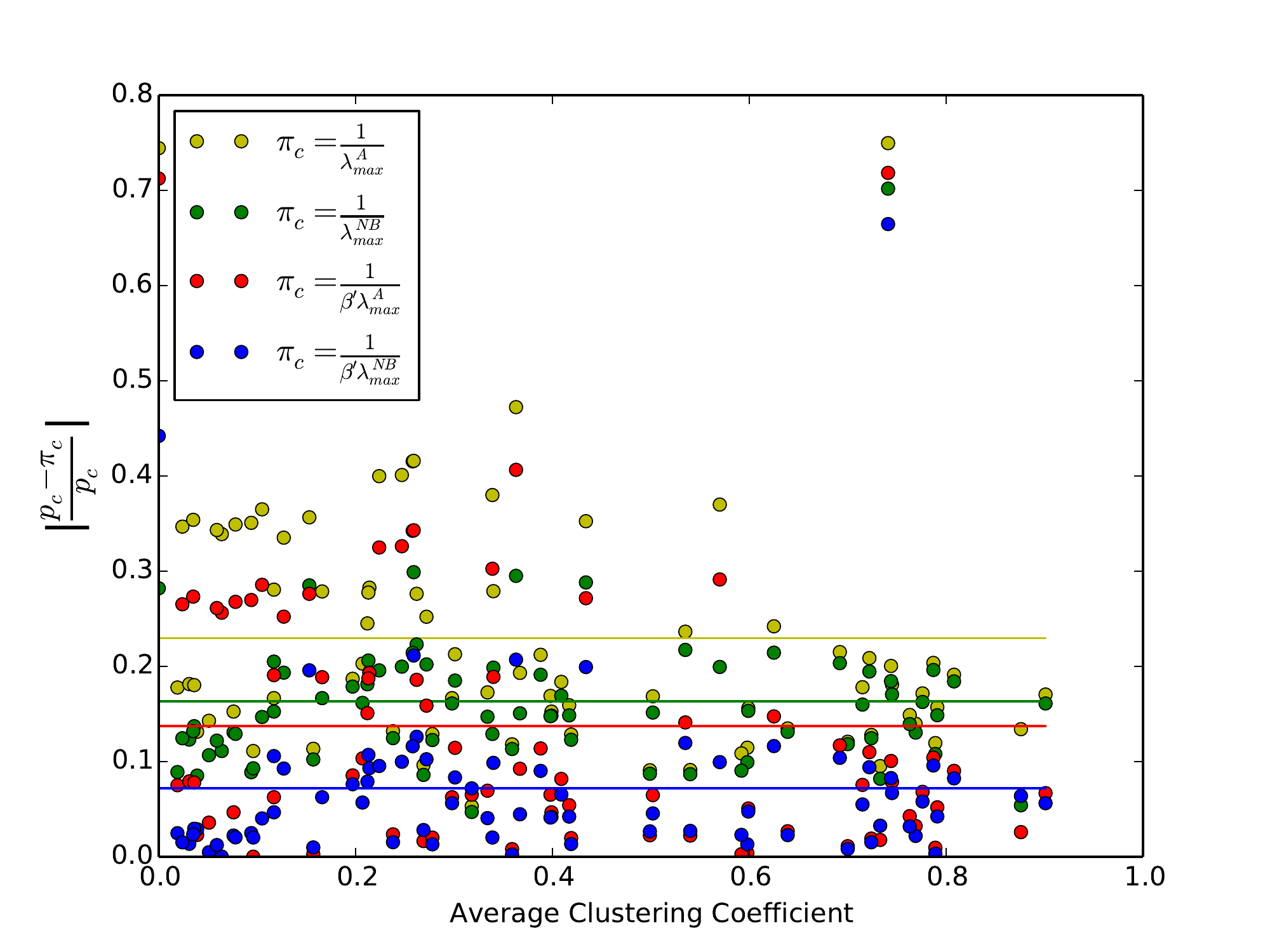}
\caption{Relative error associated to different estimates for the percolation threshold. Solid lines indicates
average values of dots of the same color.}\label{fig:est}
\end{figure}

To show that this is the case, in Figure (\ref{fig:est}) we compare different estimates of the numerical percolation threshold, 
obtained as either the leading eigenvalues of the adjacency matrix $\lambda_{max}^{A}$ or of the non-backtracking matrix $\lambda_{max}^{NB}$, 
eventually corrected by the $\beta'$ factor. 
We indeed see that $\beta'$ can be used to improve, on average, the approximation given by theoretical models. 

We believe that the corrective factor is related to finite size effects and non treelike structures, however this hypothesis needs further investigation. 
Overall, while our approach is just at infant stage and our findings are only preliminary, they may have important concrete applications. 


\subsection*{Acknowledgments}
This work was supported by the EU projects CoeGSS (grant n. 676547) and SoBigData (grant n. 654024).

\subsection*{Appendix: definition and properties of the complement network}

Formally, let $a_{ij}$ be the generic element of the adjacency matrix $\mathsf{A}$ associated with a given binary undirected graph $G$ of $N$ vertices, 
such that $a_{ij}=1$ if an edge between vertices $i$ and $j$ exists, and $a_{ij}=0$ otherwise. 
The adjacency matrix of the complement graph $\bar{G}$ is defined through $\bar{a}_{ij}=1-\delta_{ij}-a_{ij}$, 
where $\delta_{ij}$ is the Kronecker delta which excludes self loops from $\bar{G}$, and $1-\delta_{ij}$ defines the adjacency matrix of the complete graph. 
It follows trivially that $M+\bar{M}=\binom{N}{2}$, $\rho+\bar{\rho}=1$ and $k_i+\bar{k}_i=N-1$ $\forall i$, 
where $M$, $\rho$ and $k_i$ denote the number of edges, the edge density, and the degree of (number of edges incident with) generic vertex $i$, respectively. 
Thus, given the degree distribution $P(k)$, the distribution of the complement degree is obtained as $\bar{P}(\bar{k})=P(N-1-\bar{k})$, 
\ie, as the reflection of $P(k)$ on the $\frac{N-1}{2}$ vertical axis. 
Notably, the degree distribution of both a regular graph and an Erd\"os-R\'enyi graph (ER) are invariant under this transformation: 
the complement of a regular graph is a regular graph, as the complement of an ER is an ER. 
In particular, the complement of an ER with connection probability $f$ is an ER with connection probability $1-f$. 

Moving to higher-order properties, the number of triangles (closed loop of length 3) of a graph and of its complement is
\begin{equation}\label{eqn:trans_comp}
\Sigma_{\triangle} = \frac{\text{Tr}\mathsf{A}^3+\text{Tr}\mathsf{\bar{A}}^3}{6}=\binom{N}{3}-\frac{1}{2}\sum_i k_i\bar{k}_i.
\end{equation}
As such, both cases $k_i=0$ and $\bar{k}_i=0$ $\forall i$ (empty and complete graph) lead to $\Sigma_{\triangle}=\binom{N}{3}$ as expected.
As for transitivity, a complementarity relation can be written also for the local clustering coefficient $c_i= \frac{\sum_{j\neq i}\sum_{k\neq i,j}a_{ij}a_{jk}a_{ik}}{k_i(k_i-1)}$:
\begin{equation}\label{eqn:local_clust_comp}
c_ik_i(k_i -1) + \bar{c}_i \bar{k}_i (\bar{k}_i -1) = k_i^{nn}k_i + \bar{k}_i^{nn}\bar{k}_i - k_i -\bar{k}_i - k_i\bar{k}_i,
\end{equation}
where $k_i^{nn}= \frac{\sum_{j\neq i } a_{ij}k_j}{k_i}$ is the average nearest-neighbors degree.

%

\end{document}